\documentclass[a4paper, 10pt, twocolumn]{article}

\usepackage[cm]{fullpage}

\usepackage{cite}
\usepackage{lscape}
\usepackage{amsmath,amssymb,amsfonts}

\usepackage{algorithmic}
\usepackage{textcomp}

\usepackage{graphicx}
\graphicspath{{figures/}}

\usepackage{booktabs}
\usepackage{makecell}

\usepackage{textcomp}

\usepackage{url}
\usepackage{hyperref}
\usepackage{xcolor}
\usepackage{pifont}
\newcommand{\cmark}[1]{\ding{51}}

\newcommand{\subsubsubsection}[1]{\textbf{#1 --}}

\usepackage{authblk}

\begin{document}

\title{A system for objectively measuring behavior and the environment to
  support large-scale studies on childhood obesity}

\author[1, 2]{Vasileios Papapanagiotou}
\author[1]{Ioannis Sarafis}
\author[1]{Leonidas Alagialoglou}
\author[1, 3]{Vasileios Gkolemis}
\author[1, 3]{Christos Diou}
\author[1]{Anastasios Delopoulos}

\affil[1]{Multimedia Understanding Group, Department of Electrical and Computer
  Engineering, Aristotle University of Thessaloniki, 54636 Greece}

\affil[2]{IMPACT research group, Department of Medicine, Huddinge, Karolinska
  Institutet, Stockholm, Sweden}

\affil[3]{Department of Informatics and Telematics, Harokopio University of
  Athens, Greece}

\date{}

\maketitle

\begin{abstract}
  Advances in IoT technologies combined with new algorithms have enabled the
  collection and processing of high-rate multi-source data streams that quantify
  human behavior in a fine-grained level and can lead to deeper insights on
  individual behaviors as well as on the interplay between behaviors and the
  environment. In this paper, we present an integrated system that collects and
  extracts multiple behavioral and environmental indicators, aiming at improving
  public health policies for tackling obesity. Data collection takes place using
  passive methods based on smartphone and smartwatch applications that require
  minimal interaction with the user. Our goal is to present a detailed account
  of the design principles, the implementation processes, and the evaluation of
  integrated algorithms, especially given the challenges we faced, in particular
  (a) integrating multiple technologies, algorithms, and components under a
  single, unified system, and (b) large scale (big data) requirements. We also
  present evaluation results of the algorithms on datasets (public for most
  cases) such as an absolute error of $8$-$9$ steps when counting steps, $0.86$
  F1-score for detecting visited locations, and an error of less than $12$ mins
  for gross sleep time. Finally, we also briefly present studies that have been
  materialized using our system, thus demonstrating its potential value to
  public authorities and individual researchers.
\end{abstract}


\textbf{Tags:} Behavior Monitoring, Behavioral Studies, Health Informatics,
Machine Learning, Obesity, Public Health, Wearable Health Monitoring, Big Data,
Real-World Data

\section{Introduction}
\label{sec:introduction}

Obesity is recognized as a major public health challenge worldwide and its
prevalence has been rising since 1975 \cite{who2021obesity}. It affects both
physical and mental health and is linked with a range of comorbidities with
negative health outcomes. It is also a disease that is costly, severely
impacting healthcare systems throughout the world \cite{withrow2011economic}.

At the policy level, addressing obesity has proven to be challenging since its
complexity \cite{finegood2010} implies that single-element, blanket
interventions are ineffective \cite{rutter2017need}. Many experts now agree that
targeted, multi-factorial interventions are needed instead
\cite{rutter2012single}. Furthermore, prevention of obesity is easier than
treatment, and therefore interventions aiming at preventing obesity for
children and adolescents are a priority \cite{jones2011importance}.

Design of such multi-level interventions at a population level requires detailed
information about the behavior of age subgroups (children, adolescents, etc.),
as well as about the presence of obesogenic factors in their
environment. Passive collection of behavioral information through smartphones
and smartwatches (based on the sensors commonly embedded on such devices) is an
appealing alternative to questionnaire-based research since it is (i) scalable,
with volunteers using their own devices for data collection, (ii) detailed,
since information is collected continuously and with high frequency, and (iii)
objective, as behavioral information is quantified by algorithms based on sensor
signals. Furthermore, collection of environmental obesogenic factors can also be
facilitated by taking advantage of highly-detailed public repositories of census
data, public maps, etc. combined with novel, state-of-the-art algorithms to
extract obesity-related factors. Some examples of our earlier work on these
topics can be found in \cite{diou2019JIAOS} and \cite{jimaging4110125}.

To implement this approach, several technical challenges need to be
addressed. These include selecting suitable raw data types, extracting useful
and meaningful indicators accurately and/or with minimal error, handling missing
data and data quality issues, handling big data processing requirements, as well
as issues related to privacy protection. It is worth noting that most of these
challenges are not specific to obesity; they apply to most systems that perform
large-scale collection of behavioral and environmental data and indicators.

In this paper, we present the architecture of a system that enables both passive
and active collection of behavior and environment data at large scale. The
proposed system was originally designed and implemented in the context of the
EU-funded project BigO \cite{delopoulos2019, diou2020bigo} to support
acquisition of behavioral data from over $4,000$ children and adolescents in
four European countries. We discuss key requirements and design choices that
benefit any similar systems, the system architecture, along with implementation
and deployment details, extending our previous work of
\cite{papapanagiotou2020a} where we present and evaluate algorithms that are
used in the system and extract behavioral indicators. Furthermore, we briefly
outline examples of usage of this system in the school and clinical setting as
well as the value of this approach for use in official statistics.

While the proposed system heavily relies on existing technologies and
algorithms, there are some important challenges (relevant to ) that it must overcome:
\begin{enumerate}
\item There is no predetermined set of capturing apparatus. Instead, any
  smartphone and Android smartwatch can contribute data. This introduces
  multiple challenges both in creating a robust data-capturing application that
  handles all idiosyncrasies of each operating system and device vendor, as well
  as in designing and implementing pre-processing steps that curate and
  standardize all collected signals.
\item Real-world data collection happens outside of controlled experimental
  environments without any standardized protocol or script, introducing missing
  data, data with limited duration or sparse over a time-span.
\item While the application of algorithms that extract indicators is a central
  part of the system, additional algorithms are needed for aggregation and
  privacy-preserving purposes.
\item Aligning and combining data from different sources (such as individual
  behavioral indicators and environmental indicators) is a necessary,
  non-trivial task, requiring both spatial and temporal processing.
\end{enumerate}

All these challenges create a complex system that goes way beyond the
application and evaluation of individually developed algorithms under controlled
data-collection trials. The result is a single, integrated system that can
handle large volumes of data, enabling users such as public-health authorities
and individual researchers to understand obesity and its factors, and to
eventually design effective policies and solutions against obesity.

The following Section discusses relevant literature, while Section
\ref{sec:system-architecture} outlines the system architecture at component
level. Section \ref{sec:ipl} describes the data-acquisition tools and methods,
the indicator extraction algorithms, the types of indicators, and the types of
analyses. Section \ref{sec:experiments} presents examples of the evaluation
methodology along with results for various indicator-extraction
components. Section \ref{sec:real_world} presents examples of real-world uses of
the system. Finally, Section \ref{sec:conclusions} concludes the paper.

\section{Related work}
\label{sec:related_work}

Using big data in the context of obesity is generally considered to have
significant potential in facilitating research on multiple obesity-related risk
factors and their combinations \cite{morris2018can}. A recent Delphi study has
aggregated views of experts on the use of big data to tackle obesity
\cite{Vogel2019}. The study concluded that there is a need to develop (and also
encourage the adoption of) tools for reporting, ethics, data governance, and
sharing of data. Another Delphi study, carried in the context of the BigO
project \cite{o2020establishing}, shows that there is consensus among policy
makers on measuring and monitoring built, dietary and social environment
indicators with respect to obesity prevention. 

In \cite{PONTIN2021114235}, authors present an analysis of physical activity
estimation using a commercial application for smartphones in interactions with
socio-demographics. Among the main findings is that smartphone ownership (and
thus, the ability to collect big behavioral data) is not a limiting factor for
reaching lower socioeconomic groups.

Three different case studies are summarized in \cite{Wilkins2020} where big data
have been used in obesity research. In these studies, big data have been used to
associate (a) physical activity with green spaces and exercise facilities, (b)
obesity surgery and the risk of developing cancer, and (c) various
socio-demographic parameters of the Seattle area in the USA, with obesity.

Crowd-sourced data from GPS-enabled smartphones, specifically from Strava Metro
\cite{strava}, were collected during 2013-2014 in Queensland, Australia
\cite{Heesch2016}, and analyzed offline. The data are used to evaluate the
impact of changes in infrastructure on cycling behavior. Authors state that the
``passive'' method they use to collect data leads to higher volume and more
accurate analysis compared to recruitment or questionnaire-based studies.

Similarly, in \cite{OKSANEN2015135} authors perform an analysis of trajectories
of cyclists and produced heat-maps that can be used in applications such as city
planning. Data were collected from the city of Helsinki. Authors have considered
privacy and introduce $k$-anonymity to the analysis results (where both the
density of trajectories and users are taken into account).

The DAPHNE project \cite{gibbons2016data} is an EU-funded study that collected
data from an activity sensor, offering cloud-based data-analysis services. The
goal of the project is to reduce sedentary behavior, especially in the context
of obesity.

Finally, in \cite{Dontje2019-dd}, the Raine Study studies behaviors and the
environment, as well as genetics to understand social outcomes and to improve
health and well-being.

The presented architecture and system takes into consideration the findings of
these previous studies and provides implementation details for a system that can
be deployed on user-owned smartphones to facilitate data collection and
decision-support. While we tailor the system to the BigO project requirements
for obesity prevention at the policy level, we follow a generalizable
methodology suitable to any kind of behavior-monitoring system.

\section{System architecture}
\label{sec:system-architecture}

\subsection{Design considerations}
\label{sec:design-considerations}

The main objective of any such a system is to characterize population behavior
and environmental factors for each geographic segment and monitor
population-behavior changes over time as the result of policy
interventions. This is a three-step process: define the geographic segments,
extract population behavior, and extract environmental factors. The design of
such a system has to take into account the following requirements and
considerations:
\begin{enumerate}
\item ability to collect, store, and efficiently retrieve large volumes of
  high-rate data (multiple samples of multi-channel signals) from user
  smartphones to the system server(s)
\item ability to collect and cache environmental and census data from public
  repositories
\item ability to extract behavioral information from the collected signals
\item ability to aggregate, summarize, and combine the computed behavioral
  information with the environmental data and present informative views
\end{enumerate}

\begin{figure}
  \centering
  \includegraphics[width=\columnwidth]{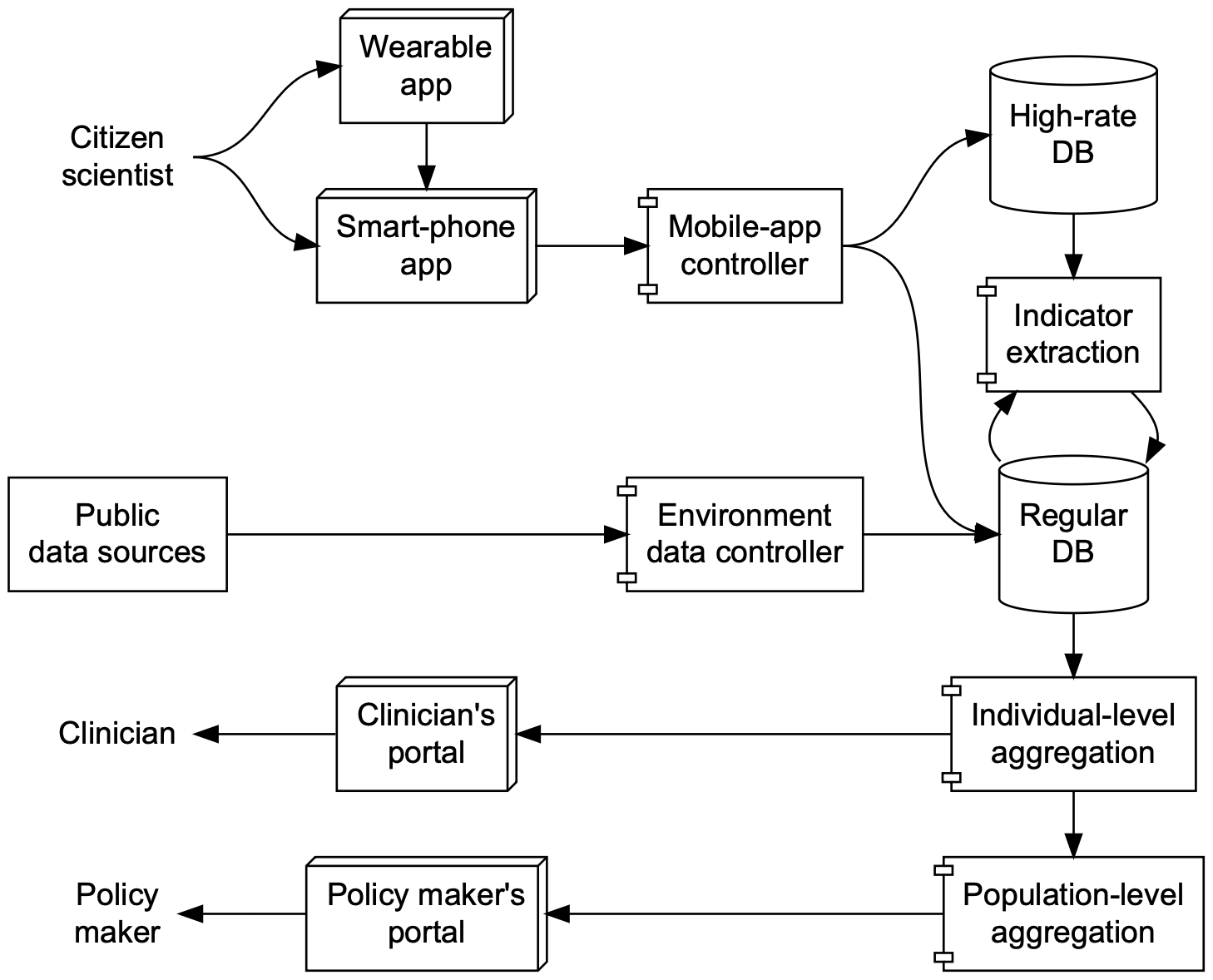}
  \caption{Generic architecture of the proposed system}
  \label{fig:abstract--arch}
\end{figure}

Based on these requirements, a generic system architecture is shown in Figure
\ref{fig:abstract--arch}.

While the above requirements can be implemented in multiple ways, the following
limitations should be taken into account:
\begin{enumerate}
\item The smartphone and wearable apps should have minimal impact on battery.
\item The volume of data transmitted between the devices and uploaded to the
  server should be minimal (for reducing the impact on both battery drain and
  bandwidth use).
\item Metadata should also be collected and used to support the detection of
  missing data (vs. periods of inactivity), to determine appropriate imputation
  methods, and to create logs and statistics.
\item Indicator extraction algorithms should be efficient, effective, and robust.
\item Population-level aggregations should be privacy preserving, i.e., they
  should not reveal information for specific individuals.
\end{enumerate}

\subsection{Design process}
\label{sec:design-process}

Several rounds of interviews and discussions with school personnel, clinicians,
public-health representatives and childhood-obesity experts led to a set of key
requirements for the proposed system \cite{info:doi/10.2196/14778}. These
include both functional requirements related to the types of measurements
recorded by the system, as well as non-functional requirements related to user
privacy.

The users of the system are: (a) participating children and their teachers, to
support educational activities on obesity prevention, (b) clinicians, who use
aggregated measurements of their patient behavior to offer personalized
consultation, and (c) public-health policy makers, who view measurements of
population behavior and their environment as evidence to support policy
decisions and monitor changes to evaluate the effectiveness of applied forces.

Behavioral data are primarily collected by the participants through their
smartphone and smartwatch accelerometer and GPS sensors. Pictures are also
contributed to self-report meals and food advertisements. In this context, the
proposed system facilitates a participatory-sensing approach by enabling
children and adolescents to provide data that help to better understand the
local obesogenic environment. Environmental data is collected from publicly
available repositories, including statistical authorities, census data sources,
and public maps.

\begin{figure}
  \centering
  \includegraphics[width=.5\textwidth]{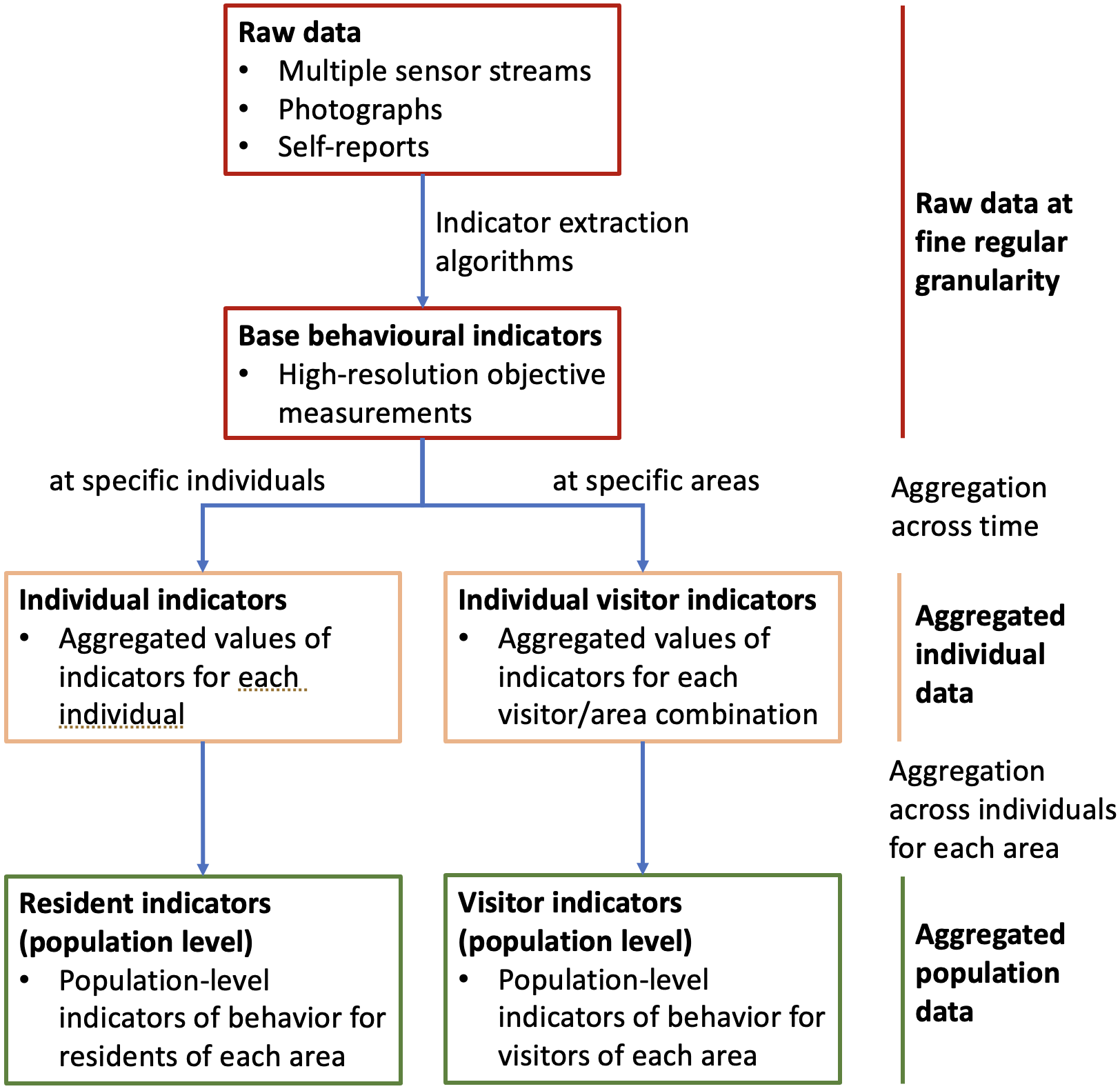}
  \caption[Data aggregation]{Data aggregation: from raw data to population-level
    behavioral indicators}
  \label{fig:data-aggregation}
\end{figure}

The collected data is processed at three layers (Figure
\ref{fig:data-aggregation}).

The first layer is responsible for (a) data acquisition from smartphone and
smartwatch sensors and external sources, and (b) extracting behavioral
indicators (i.e., well-defined and quantifiable pieces of information) at a high
temporal granularity (e.g., one value per minute).

At the second layer, behavior indicators such as ``number of steps'' and
``visits to fast-food restaurant'' are aggregated (across time) at two levels:
(a) at the level of an individual, to summarize his/her behavior, and (b) at the
level of geofencing (i.e., geographic segments), to summarize behavior that is
exhibited within the limits of predefined geographical borders, typically
covering the area of a neighborhood.

Finally, in the third layer, data are aggregated across the population,
separately for each of the individual-based and geofence-based aggregations of
the previous layer. This information is used for statistical analysis and to
produce descriptive statistics, summary views, and choropleth maps.

\begin{figure*}
  \centering
  \includegraphics[width=.8\linewidth]{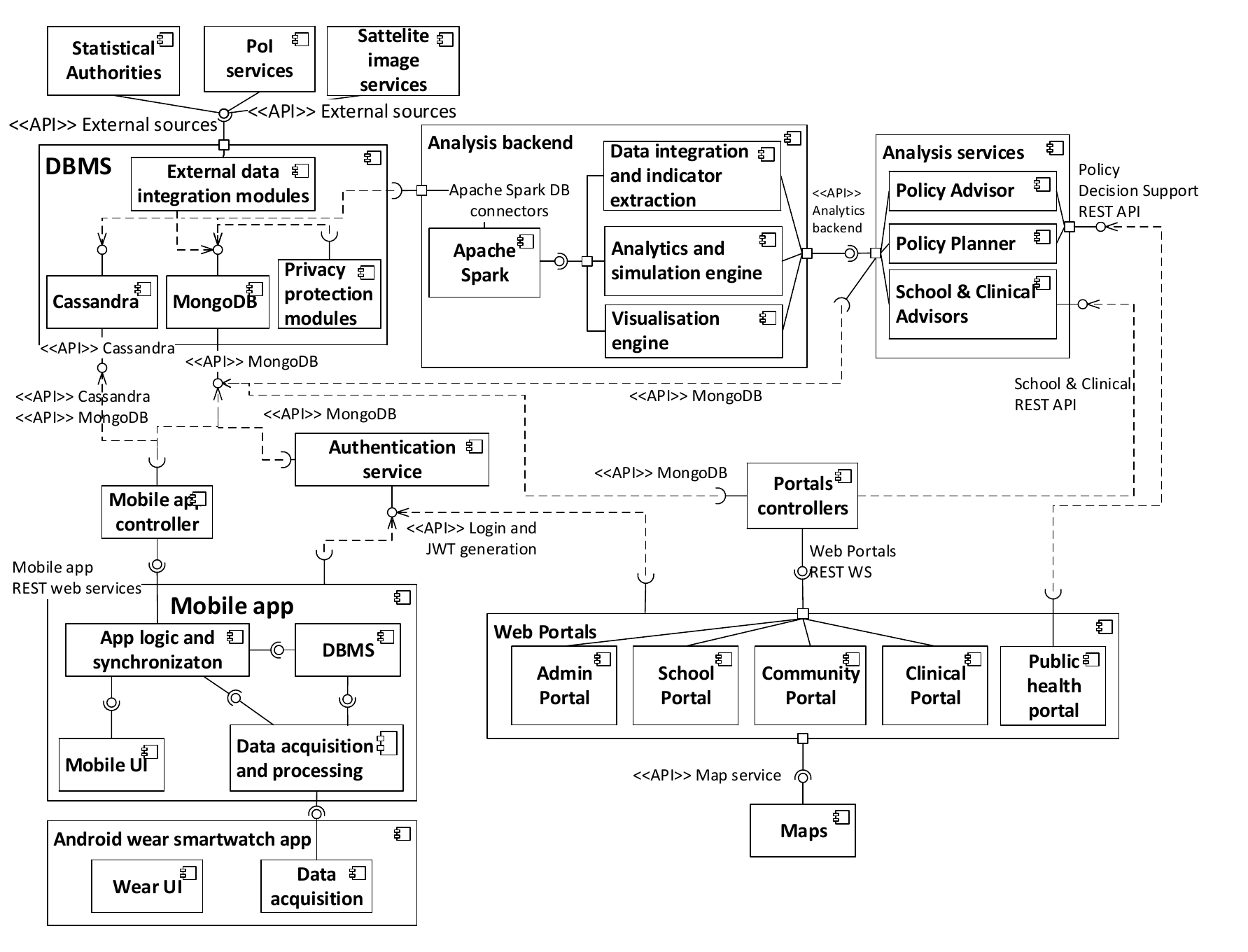}
  \caption{UML Component diagram of the proposed system architecture.}
  \label{fig:architecture}
\end{figure*}

To support these functionalities and satisfy the requirements and
considerations of Section \ref{sec:design-considerations}, the system is
designed as shown in the UML component diagram (Figure
\ref{fig:architecture}). 

The smartphone and Android-Wear smartwatch apps are designed with separate user
interface (UI) and data-acquisition components, while the smartphone app also
includes components for synchronization with the Android wear app, a local (to
the smartphone) database for storing the collected data, metadata, and other
diagnostic information, along with a server-uploading component. The Web Portals
are independent web applications which also communicate with external map
services.

Both the smartphone app and the web portals communicate with the server-side
modules only through dedicated controller components (the mobile app controller
and the portals controller, Figure \ref{fig:architecture}), and a single
authentication server is used to control access.

The server-side modules are organized in three groups. The database management
system (DBMS) group is responsible for storing the high volumes of raw data that
are collected by the smartphone app as well as communicating with external
sources of environmental data (i.e., publicly available maps, census data, and
demographic data sources). The analysis back-end modules process the collected
data into meaningful behavioral and environmental indicators and perform
aggregations. Finally, the analysis services implement the functionalities
required for the web portals.

\section{Information processing layers}
\label{sec:ipl}


\subsection{Extracting population behavior}
\label{sec:extracting-population-behaviour}

Extracting population behavior measurements requires computation of multiple
behavioral indicators from the signals collected by the smartphone and
smartwatch. We identify three main groups of such indicators: physical activity,
location and transportation, and sleep.

\subsubsubsection{Physical activity} We compute activity counts, steps, activity
level (i.e., sedentary, moderate, vigorous, and very vigorous), and activity
type (e.g., standing, walking, running) at regular intervals (i.e., once per
minute).  The input to all physical-activity algorithms presented in this
section is the 3-dimensional (3D) accelerometer signal, at a relatively low
sampling rate (usually in the range of $5$ to $25$ Hz, depending on the device).

It should be noted that modern Android devices that are supported by our system
offer different levels of sampling rates (``slow'' \cite{androidsampling0},
``medium'' \cite{androidsampling1}, ``fast'' \cite{androidsampling2}, and
``maximum'' \cite{androidsampling3}) , however these correspond to sampling
rates that vary between different vendors and devices (i.e., smartphones and
smartwatches). Initially, we developed part of the system using the ``medium''
sampling rates, which usually correspond to values around $50$ Hz. However, this
caused significant battery drain and the first users of the system had multiple
complaints regarding this issue, leading to low user acceptance. Thus, we
switched to the ``slow'' sampling rate (which is between $5$ to $25$ Hz) and
adjusted our algorithms accordingly. While this choice did reduce the quality of
extracted indicators, it was necessary to maintain an acceptable user
experience.

Activity counts are commonly used in the physical-activity literature and are
typically computed by devices manufactured by Actigraph \cite{actigraph}. These
devices have been used extensively both in sleep-related \cite{Ancoli2003} and
physical activity-related studies \cite{Trost2011}.

Counting steps during walking is one of the most common indicators measured by
many devices, i.e., fitness trackers (by companies like Fitbit or Garmin),
smartwatches (Apple watch, Android smartwatches, etc.), and even smartphones. In
most cases, proprietary algorithms and methods are used to estimate the number
of walked steps, and sometimes these methods are integrated into the hardware,
and are provided as ``pedometer'' sensors. To avoid discrepancies and
differences between different software and hardware, two state-of-the-art
algorithms are implemented in BigO, in particular the algorithms of
\cite{gu2017robust} and \cite{genovese2017smartwatch}. The algorithm
\cite{gu2017robust} is used on smartphone devices and is reported to achieve an
error of $1\%$ on free walking and $11.67\%$ false walking on a dataset of eight
volunteers performing $300$ steps each on two trials (ground truth was created
by volunteers actively counting their steps as they walked). The algorithm of
\cite{genovese2017smartwatch} is used on smartwatch devices; it achieves a
relative error of $1\%$ to $3\%$ on a dataset of eight people repeating $80$
trials (ground truth was created by an external observer who counted the
steps). Different algorithms are used for the two types of devices since the
nature of the captured acceleration signals are different in each case. In
particular, the algorithm of \cite{gu2017robust} mainly focuses on cases where
the accelerometer is in the subject’s pocket or back-pack while the algorithm of
\cite{genovese2017smartwatch} specifically focuses on smartwatches. Using the
appropriate algorithm for each device type helps increase the effectiveness and
reduce the overall absolute error between the detected and actual (ground truth)
steps.

Detecting activity type involves processing the accelerometer signal to
determine which activity the user is performing, from a predefined list of
activity types. Detection is achieved using a machine-learning model, built for
example using support vector machines (SVM), which can be trained on datasets
with a predefined set of target classes. Common activity types include sitting,
standing, walking, jogging, running, ascending stairs, and descending stairs;
sometimes more fine-grained types are included, such as dish washing, vacuum
cleaning, playing football, playing basketball, etc.

The approach of \cite{reiss2012creating} has been implemented for activity-type
recognition, with some important differences between the application of the
algorithm in the original paper and in BigO. These are:
\begin{itemize}
\item Accelerometer is sampled at a higher rate in \cite{reiss2012creating}
\item Multiple accelerometers are used simultaneously in
  \cite{reiss2012creating} (their windows are synchronized in time and their
  feature vectors are concatenated); in our work, we only use one stream
\item A heart-rate sensor is also used in \cite{reiss2012creating}; in our work,
  we do not use the heart-rate sensor for physical activity
\end{itemize}

The limitations applied in BigO reduce the effectiveness of the algorithm,
however the reduction is small (see Section \ref{sec:experiments}) while it
enables the practical application of the algorithm, i.e., does not require a
user to wear a smartwatch on each hand, or for the smartphone to be strapped on
belt, and does not require a constant heart-rate unit.

\begin{figure}
  \centering
  \includegraphics[width=.999\columnwidth]{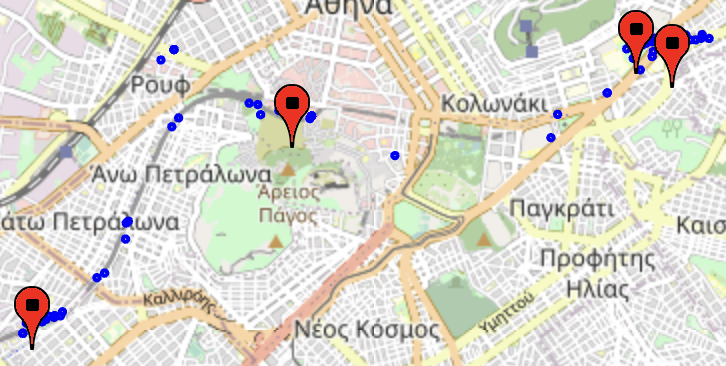}
  \caption{Point-of-interest detection example with development data. Map data
    from Open Street Map.}
  \label{fig:location-example}
\end{figure}

\subsubsubsection{Location and transportation} Indicators related to location
and transportation are computed on an event-based scale (i.e., they depend on
when an individual arrives and departs from a specific location).  In
particular, we identify visited points-of-interest (including restaurants, gyms,
parks, supermarkets, etc.), some locations with special significance (home and
school), and also when and how the user moves from each location to another,
i.e. transportation mode (e.g. walking, bicycling, using a bus). The input to
all algorithms in this section is a sequence of time-stamped location
coordinates (latitude and longitude). Additionally, the transportation-mode
detection algorithm also uses the acceleration signal (this is the same signal
that is used for extracting the physical-activity indicators). Figure
\ref{fig:location-example} illustrates examples of detected PoIs and
transportation mode, using publicly available datasets and development data
collected within BigO.

To detect PoIs, we follow the approach of \cite{luo2017improved} that detects
visited locations from a set of recorded coordinates (called location points
from now on). The algorithm introduces a metric called move-ability and computes
a density value that is used by a modified version of the DBSCAN algorithm.

Home and school detection relies on detecting the most frequently-visited
PoIs. We apply heuristic rules that take into account frequency, time of day,
and weekday, and recognize which PoI is the user's home and which one is their
school.

Note that once the type (e.g., park, restaurant) of a visited PoI has been
determined, the actual coordinates of the visited PoI are deleted. This is done
for data minimization and privacy protection purposes.

Detecting transportation mode requires detecting visited PoIs first, since a
subject cannot be traveling while visiting a PoI. Thus, time intervals between
PoIs can be examined as candidate ``trips''. Once a time interval is identified
as a trip, the corresponding accelerometer data is extracted and used to detect
transportation mode based on the algorithm of \cite{shafique2016travel}. The
implementation of BigO has some minor differences from the algorithm as
presented in \cite{shafique2016travel}, in particular:
\begin{itemize}
\item Use of smaller windows ($1$ sec instead of $10$ sec) to allow more
  fine-grained detection
\item Additional features (power spectral density)
\item Different classifier (SVM instead of random forest)
\item A post-processing median filter
\end{itemize}

\subsubsubsection{Sleep} For estimating sleep behavior, we use the acceleration
signal we collect from the smartwatch. While there are a lot of from
dominant-hand worn sensors \cite{doherty2017large}, in our case the smartwatch
was usually worn on the wrist of the non-dominant hand.  Although a lot of
methods have been implemented recently for estimating sleep from raw
accelerometer data, they are not drastically different from older methods, nor
are they extensively evaluated \cite{tilmanne2009, vanhees2015, galland2012}. On
the other hand, initial methods based on activity counts have been tested in
multiple scenarios both in research studies and in the industry. Among them, the
most famous are \cite{cole1992} and \cite{sadeh1994}. These methods have been
evaluated in a wide range of studies \cite{galland2012} and have demonstrated
satisfactory results. For this reason, we choose these methods and adapt their
implementation for our proposed system (i.e., using different sampling rates).

Both implementations are based on activity counts extracted from $1$-minute
windows; they output boolean predictions (sleeping vs. not-sleeping) for each
$1$-minute window. Based on these boolean labels, we compute six indicators:
gross sleep time (GST), sleep start (SS), sleep end (SE), total time of
interrupts (TTI), number of interrupts (NI), and net sleep time (NST).

\subsection{Spatial aggregation}

To enable the collection of behavioral information for locations without
compromising the anonymity of the users, we employ a mechanism that aggregates
behavior across spatial regions. This mechanism is based on the ``geohash
system'' \cite{geohash, morton1966computer}, which is a public-domain geocoding
mechanism. The geohash system encodes GPS coordinates to short, alphanumerical
strings; each string (a ``geohash'') is a unique identifier for a rectangular
geographic area. This is the main mechanism for encoding geographical regions in
the proposed system.

An advantage of using geohashes is that it is possible to control the
granularity of the area, simply by removing characters from the end of the
geohash.

The ability to easily apply various levels of privacy protection regarding an
individual’s location makes the geohash system appropriate for collecting
spatial behavioral measurements. In BigO, this data collection is realized
through the geohash votes mechanism, which operates as follows:
\begin{enumerate}
\item The system detects a visit of an individual inside a particular geohash
\item The system measures the individual’s behavior during the visit (details
  are provided in Section \ref{sec:extracting-population-behaviour})
\item A anonymous vote is cast, indicating the behavior of one individual
  during his/her stay at that geohash
\end{enumerate}

Given enough users and time, each geohash (within a study area, e.g., a city or
municipality) is populated with many votes from multiple users. It is important
to note that these votes are anonymous, i.e., they cannot be tracked back to the
specific user who cast the vote (the system, however, does keep track of a
mapping using anonymized IDs to avoid double voting). The votes of each geohash
can then aggregated in many ways and summarize behavior of the population within
very specific geographical limits.

Using the geohash-votes mechanism, we can collect aggregated information about
the behavior of the population during their visits at particular locations,
without the need to store individual location data centrally. An alternative
(but similar) representation can result if individuals cast votes in fixed,
regular time intervals (e.g., every quarter, or every hour). This option,
however, can conceal the total visit time of a subject to a region (geohash).

\subsubsection{Aggregation and data analysis}

The behavioral indicators that are computed by the algorithms of Section
\ref{sec:extracting-population-behaviour} have high temporal resolution, and
correspond to individual events (e.g., a visit to a restaurant, a morning walk
to school) or to measurements with high sampling rate (e.g., activity type
performed every minute). We call these ``Base indicators''. The detailed
information of base indicators cannot be easily visualized or communicated for
the purposes of behavioral research, or to support policy
decisions. Additionally, visualizing base behavioral indicators or using them
directly for analysis may raise privacy concerns.

The proposed system therefore supports data aggregation functionalities that
produce measurements summarizing individual or population behavior, as in
\cite{diou2019JIAOS}. Aggregation can take place at the individual and at the
population level.

\subsubsubsection{Individual-level aggregation} The supported aggregation
functions include summation, averaging, and distribution estimation using
histograms. Applying these aggregations to base indicators results in
\begin{itemize}
\item \emph{Individual visitor indicators:} aggregation of individual behavior
  during the time spent at a specific region (geohash). Examples include the
  average number of steps per hour during visits to the location, or the
  percentage of visits to the location that resulted in a visit to a cafe.
\item \emph{Individual resident indicators:} aggregation of individual behavior,
  irrespective of location. In this case indicators are computed by aggregating
  the overall behavior of individuals that reside in a specific
  geohash. Examples include the time spent performing the ``running'' activity
  per week, the distribution of monthly visits to a set of location types (e.g.,
  gym, restaurant, park). "Behavior profiles" can be used as a formal way to
  describe an individual’s behavior and as input to machine learning and
  analysis algorithms \cite{8857161}. A behavior profile first aggregates data
  of PoIs and transitions to generate a probabilistic mobility graph for each
  user. In addition, aggregated indicators are calculated for each node (i.e.,
  PoI type), such as physical activity and time spent, and for each edge (i.e.,
  transition), such as distance and transportation mode.
\end{itemize}

Note that spatial aggregation can be implemented in both cases. In the first
case, the vote is cast to the visited geohash, while in the second case the vote
is cast to the user's resident geohash (even if the activity took place
elsewhere).

One difficulty that occurs when applying aggregation functions is handling
missing data. It is common to have periods with missing recordings for reasons
that include (a) sensor or application failure, (b) the recording service being
stopped by the smartphone operating system and (c) the user manually stopping
data acquisition. The approach that we followed in this system implementation
is to apply a filter to remove days with recording time below a threshold from
the aggregation. Moreover, using normalized measures, such as average number of
steps per hour instead of absolute number of steps can make indicators from
users with different amount of data recordings comparable.

\subsubsubsection{Population-level indicators} The next step is to aggregate
individual-level measurements to produce population-level behavioral
indicators. Given a neighborhood (defined as a geographical region) or geohash,
we can again distinguish between \emph{visitor} and \emph{resident} indicators.

Visitor indicators summarize the behavior of individuals during their visit in
the neighborhood. These can be computed directly by applying an aggregation
function to the individual visitor indicators of the previous section. Resident
indicators, on the other hand, summarize the behavior of individuals living in
a particular neighborhood, irrespective of where their behavior took
place. These can be computed by aggregating individual indicators.

Examples of visitor indicators include and ``percentage of neighborhood visitors
that performed at least 10 minutes of the running activity'', ``average number
of steps per hour of visitors'' and ``average time spent in restaurants per
visitor''. Examples of resident indicators include the ``daily average number of
steps'', ``number of weekly sessions of at least 20 minutes of walking'' and
``average number of weekly visits to restaurants''.

Furthermore, the geohash-votes mechanism has been used in combination with local
environmental characteristics to produce machine learning models for estimating
population behavior in new areas. This approach has demonstrated high predictive
performance \cite{9176246} and can be used for generalizing in areas without
data (e.g., different parts of the same metropolitan area where population data
are not available) or for imputation purposes when visualizing heatmaps (e.g.,
geohashes that happen to have no votes or very few votes that fall below the
privacy thresholds).

To avoid introducing bias, only individuals with sufficient volume of data are
considered in the computation of the population-level indicators.

\subsection{Extracting environmental factors}
\label{sec:extracting-environmental-factors}

While the chosen geographic segment (in the context of BigO) is the geohash,
environmental information is usually available in different scales such as an
administrative region. Given the focus of BigO on providing decision support for
public health authorities, this information plays a central role for analysis.

The main driver for the selection of environmental factors to be modeled in BigO
has been the available literature and the Delphi study
\cite{o2020establishing}. In total, $46$ factors were identified, in the
following $5$ categories:
\begin{enumerate}
\item \emph{Urban environment} ($20$ factors), e.g., number of supermarkets and
  grocery stores, number of restaurants and food outlets
\item \emph{School} ($6$ factors), e.g., school exercise programs, school
  meals/breaks
\item \emph{Socioeconomic, pricing, and inequalities} ($10$ factors): e.g.,
  average income in neighborhood, education-level statistics
\item \emph{Food marketing} ($3$ factors), e.g., exposure to food advertisements
  from TV
\item \emph{Policy} ($7$ factors), e.g., sugar tax, salt tax
\end{enumerate}

Automatically extracted environmental factors result from information available
through statistical authorities, GIS platforms etc. Others need to be manually
measured and provided, such as the pricing of different types of goods.

Depending on the country or region, some of the factor values may not be
available. In other cases, data is available but not at the required level of
geographical detail. In these cases, one may use methodologies that consider
surrogate variables, such as the one proposed in \cite{jimaging4110125}.

\subsection{Final list of behavioral indicators}

Based on the information extracted from the algorithms described so far,
multiple population-behavior indicators can be computed, on different aggregation
levels (both temporal and spatial). Table \ref{tab:indicators} presents the
consolidated list of population-level behavioral indicators that we extract.

\begin{table*}
  \centering
  \caption{List of aggregated behavioral indicators per spatial region}
  \label{tab:indicators}
  \begin{tabular}{l|ccc|ccc}
    \toprule
    \textbf{Indicator} & \multicolumn{3}{c|}{\textbf{Aggregation per}} & \multicolumn{3}{c}{\textbf{Sensor}} \\
    & \emph{resident} & \emph{visitor} & \emph{vote} & \emph{acc.} & \emph{loc.} & \emph{G.I.S.} \\
    \midrule
    \multicolumn{6}{l}{\textit{Physical activity}} \\
    Average: & & & & & \\
    - activity counts/minute & & \cmark{7} & \cmark{8} & \cmark{} & & \\
    - hourly steps & & \cmark{5} & \cmark{6} & \cmark{} & & \\
    - daily steps & \cmark{18} & & & \cmark{} & & \\
    - sedentary minutes after school & \cmark{19?} & & & \cmark{} & \cmark{} & \\
    - sleep duration & \cmark{} & & & \cmark{} & & \\
    Distribution of physical activity: & & & & & & \\
    - types & \cmark{17} & & & \cmark{} & & \\
    - levels & & \cmark{1} & \cmark{2} & \cmark{} & & \\
    Percentage of individuals with: & & & & & & \\
    - sedentary-behavior & & \cmark{3} & \cmark{4} &  \cmark{} & & \\
    - sedentary average activity level & \cmark{16} & & & \cmark{} & & \\
    - at least $60$ min of average daily walking time & \cmark{15} & & & \cmark{} & & \\
    \midrule
    \multicolumn{7}{l}{\textit{Visits}} \\
    Percentage of votes with at least one visit to: & & & & & & \\
    - a food-related location & & & \cmark{9} & & \cmark{} & \cmark{} \\
    - to supermarkets or grocery stores & & & \cmark{10} & & \cmark{} & \cmark{} \\
    - a fast food or a takeaway restaurant & & & \cmark{11} & & \cmark{} & \cmark{} \\
    - a public park & & & \cmark{12} & & \cmark{} & \cmark{} \\
    - a recreational facility & & & \cmark{13} & & \cmark{} & \cmark{} \\
    - an athletics or sports facility & & & \cmark{14} & & \cmark{} & \cmark{} \\
    Average number of weekly visits to: & & & & & & \\
    - restaurants & \cmark{20} & & & & \cmark{} & \cmark{} \\
    - food outlets & \cmark{21} & & & & \cmark{} & \cmark{} \\
    - cafes & \cmark{22} & & & & \cmark{} & \cmark{} \\
    - fast food locations & \cmark{23} & & & & \cmark{} & \cmark{} \\
    - food-related locations & \cmark{24} & & & & \cmark{} & \cmark{} \\
    - supermarkets or grocery stores & \cmark{25} & & & & \cmark{} & \cmark{} \\
    - restaurants or food outlets & \cmark{26} & & & & \cmark{} & \cmark{} \\
    - take-away restaurants & \cmark{27} & & & & \cmark{} & \cmark{} \\
    - fast food or take-away restaurants & \cmark{28} & & & & \cmark{} & \cmark{} \\
    - bars & \cmark{29} & & & & \cmark{} & \cmark{} \\
    - wine or liquor stores & \cmark{30} & & & & \cmark{} & \cmark{} \\
    - public parks & \cmark{31} & & & & \cmark{} & \cmark{} \\
    Average number of visits to: & & & & & & \\
    - indoor recreational facilities & \cmark{32} & & & & \cmark{} & \cmark{} \\
    - athletics or sports facilities & \cmark{33} & & & & \cmark{} & \cmark{} \\
    \midrule
    \multicolumn{7}{l}{\textit{Mobility}} \\
    Distribution of transportation modes & \cmark{} & \cmark{} & \cmark{} & \cmark{} & & \\
    \bottomrule
  \end{tabular}
\end{table*}

\section{Experiments \& results with indicator extraction algorithms}
\label{sec:experiments}

This section presents the methodology for evaluating each indicator-extraction
algorithm of the system. The experiments and results presented here are not
exhaustive; they instead focus on some of the fundamental algorithms of the
system (such as the step-counting algorithm or the sleep-detection algorithm)
and present evaluation results on publicly available datasets. However, a
similar methodology can be followed for each of the behavioral indicators
presented in Table \ref{tab:indicators}, given appropriate ground truth.

\subsection{Physical activity}

Step-counting is evaluated on the publicly-available dataset of
\cite{brajdic2013}. It includes time-series recordings from $27$ participants
using smartphones; each participant performed $130$ walks using six smartphone
placements simultaneously with an average of $78\pm8.34$ steps per walk. The
recordings were made without any constraints.  The smartphones were placed at
different positions, such as front or back trouser pocket, handbag, or were held
by the participants. Additional datasets also exist \cite{mattfeld2017new,
  small2022oxwalk}, however, they were published after the project's main
development period and thus they were not available for experimentation during
that time.

The number of detected steps is validated against ground truth, and results are
presented in Table \ref{tab:step_counting}. Ground truth number of steps,
detected steps, absolute and relative errors are shown per device
position. Overall absolute error is $8$ steps (per recording), and overall
relative error is $9.7\%$.

\begin{table*}
  \centering
  \caption{Evaluation results (average error and std) of step calculation for
    the different phone placements of the dataset of \cite{brajdic2013} for the
    algorithm of \cite{gu2017robust}}
  \label{tab:step_counting}
  \begin{tabular}{lrcccc}
    \toprule
    \textbf{Position}
    & \textbf{\#}
    & \makecell{\textbf{Predicted}\\\textbf{steps}}
    & \makecell{\textbf{Absolute}\\\textbf{error}}
    & \makecell{\textbf{Relative}\\\textbf{error (\%)}}\\
    \midrule
    Hand-held
    & $21$ & $73 \pm 16$ & $8 \pm 13.4$ & $10 \pm 15.5$\\
    Hand-held \& using
    & $21$ & $74 \pm 13.6$ & $9 \pm 14.3$ & $10 \pm 14.2$\\
    \makecell[l]{Trousers\\\ \ back pocket}
    & $2$ & $82 \pm 19$ & $6 \pm 5.7$ & $7 \pm 6.3$\\
    \makecell[l]{Trousers\\\ \ front pocket}
    & $1$ & $75 \pm 0$ & $7 \pm 0.0$ & $10 \pm 0.0$\\
    Handbag
    & $5$ & $94 \pm 9.6$ & $14 \pm 5.5$ & $18 \pm 6.8$\\
    Backpack
    & $16$ & $77 \pm 7.9$ & $5 \pm 3.8$ & $6 \pm 4.7$\\
    Shirt pocket
    & $1$ & $94 \pm 0.0$ & $11 \pm 0.0$ & $13 \pm 0.0$\\
    \bottomrule
  \end{tabular}
\end{table*}

For evaluating physical activity type detection, the PAMAP2 dataset
\cite{reiss2012creating} is used. It includes $12$ activity types, both basic
every-day activities, postures, and motion types. Each participant performed all
activities in a predefined order; each activity was performed for $3$ minutes
approximately (the only exceptions that lasted less being ascending and
descending stairs and rope jumping). In total, $8$ hours of data were collected.

\begin{table*}
  \centering
  \caption{Activity type recognition classification results on the dataset of
    \cite{reiss2012creating}}
  \label{tab:activity_type}
  \begin{tabular}{l|cccccc}
    \toprule
    & \textbf{lay} & \textbf{stand} & \textbf{walk} & \textbf{run} & \textbf{cycle} & \textbf{stairs}\\
    \midrule
    \textbf{lay}     & $94.4$ &  $5.5$ &  $0.0$ &  $0.0$ &  $0.1$ &  $0.1$ \\
    \textbf{stand}   &  $0.8$ & $95.3$ &  $0.8$ &  $0.5$ &  $1.8$ &  $1.0$ \\
    \textbf{walk}    &  $0.0$ &  $1.1$ & $86.4$ &  $0.1$ &  $0.6$ & $11.8$ \\
    \textbf{run}     &  $0.0$ &  $3.3$ &  $0.4$ & $94.3$ &  $0.3$ &  $1.7$ \\
    \textbf{cycle}   &  $0.5$ &  $7.6$ &  $3.5$ &  $0.7$ & $83.6$ &  $4.0$ \\
    \textbf{stairs}  &  $0.1$ &  $3.0$ & $12.8$ &  $2.0$ &  $1.6$ & $80.5$ \\
    \bottomrule
  \end{tabular}
\end{table*}

\subsection{Location and transportation}

PoI detection and transportation mode recognition algorithms are evaluated on
the Sussex-Huawei locomotion (SHL) dataset \cite{shl1, shl2}. The dataset was
collected by the Wearable Technologies Lab at the University of Sussex as part
of a research project funded by Huawei. It contains multiple modes of locomotion
and transportation, as captured by various sensors (inertial, location, audio,
and more) of smartphones. Currently, data from $3$ participants are available,
that include $3$ days of recordings per participant.

Evaluation of PoI algorithm is performed on the SHL dataset, using the location
data as input and the ground truth state to determine visited locations.

Classification results are computed by performing a ``matching'' process between
ground-truth and detected PoIs. This process is described by the
following rules
\begin{itemize}
\item Each ground-truth PoI can be matched with at most one detected PoI
\item Each detected PoI can be matched with at most one ground-truth PoI
\item Matching requires a maximum of distance between ground-truth and detected
  center co-ordinates
\item Matching requires a minimum of overlap in time between ground-truth and
  detected PoIs
\end{itemize}

Each pair of matched ground-truth and detected PoIs is counted as $1$ true
positive (TP) detection. Each non-matched ground-truth PoI is counted as $1$
false negative (FN) detection. Finally, each non-matched detected PoI is counted
as $1$ false positive (FP) detection.

Table \ref{tab:visited_locations} presents the classification results. All TP,
FP, and FN are shown for each subject, as well as three additional metrics:
precision, recall, and F1-score. Additionally, total results are shown in the
last line of the results Table. Total results are computed by summing TP, FP,
and FN from all subjects in order to obtain TP, FP, and FN for the entire
dataset, and the remaining metrics are computed on the basis of the entire
dataset TP, FP, and FN.

\begin{table*}
  \centering
  \caption{Evaluation of the visited-locations detection algorithm
    \cite{luo2017improved} on the SHL dataset.}
  \label{tab:visited_locations}
  \begin{tabular}{lcccccc}
    \toprule
    \textbf{Subject} & \textbf{TP} & \textbf{FP} & \textbf{FN}
    & \textbf{Precision} & \textbf{Recall} & \textbf{F1-score}\\
    \midrule
    $1$ & $8$ & $0$ & $1$ & $1.00$ & $0.89$ & $0.94$\\
    $2$ & $15$ & $2$ & $4$ & $0.88$ & $0.79$ & $0.83$\\
    $3$ & $13$ & $2$ & $3$ & $0.87$ & $0.81$ & $0.84$\\
    sum & $36$ & $4$ & $8$ & $0.90$ & $0.82$ & $0.86$\\
    \bottomrule
  \end{tabular}
\end{table*}

Detection of transportation mode is evaluated on the SHL dataset. The
leave-one-subject-out scheme was used to train and test the classification
models on the three users of the dataset.

The different transportation modes that are available are walk/run, bike, car,
bus, and train/subway. In total, $7$ modes were merged into $5$ due to
similarities between train and subway in one case, and walking and running in
the other case (walking can be differentiated from running by the activity type
detection algorithm). The number of data points per class were not equal in the
dataset, so class weights were used during model training. Class weights were
chosen as $w_{i}=\min_{k}{n_{k}}/n_{}{i},\,i=1,2,3,4,5$ where $n_{i}$ is the number
of data points of the $i$-th class that are available during
training. Additionally, $C$ parameter was set equal to $1000$, while the
radial-basis-function kernel was used with parameter $\gamma=1/D$ where $D$ is the
number of features (feature-vector dimension).

\begin{table*}
  \centering
  \caption{Evaluation of the transportation-mode detection algorithm on the SHL
    dataset. Results are given per window.}
  \label{tab:transportation_mode}
  \begin{tabular}{l|ccccc}
    \toprule
    & \textbf{walk/run} & \textbf{bike} & \textbf{car} & \textbf{bus} & \textbf{train/subway} \\
    \midrule
    \textbf{walk/run}     & $2990$ & $163$ & $68$ & $119$ & $39$ \\
    \textbf{bike}         & $138$ & $1487$ & $4$ & $0$ & $0$ \\
    \textbf{car}          & $1$ & $7$ & $1690$ & $962$ & $293$ \\
    \textbf{bus}          & $6$ & $1$ & $380$ & $1087$ & $320$ \\
    \textbf{train/subway} & $1$ & $11$ & $48$ & $870$ & $5039$ \\
    \bottomrule
  \end{tabular}
\end{table*}

\subsection{Sleep}

Two datasets are being used for evaluating the performance of the sleep
detection methods. Newcastle dataset \cite{vanhees2015} which is publicly
available and a dataset we created on our own (BigO dataset). It should be noted
that both datasets contain recordings only from adults, as it was not possible
to obtain recordings from children (i.e., the BigO's target population) with
ground truth.

Newcastle dataset, provided by \cite{vanhees2015}, is the only publicly
available dataset that combines raw accelerometer data and ground truth labeling
from polysomnography test. The study took place in Newcastle upon Tyne (UK) in
2015. The data comes from a single night recording of $28$ sleep clinic
patients. The majority of them are aged over $50$ years old and face serious
sleep disorders, which makes them unsuitable for validating our algorithms. For
this reason, we isolated the unique 6 subject that do not suffer from major
sleep disorders and can be characterized as normal sleepers.

Recordings last for an average of $9.13$ hours and contain one sleep recording of
mean duration of $7.02$ hours. Total time of interrupts during sleeping periods is
$35$ minutes as average, distributed in $16$ distinct interrupts.

Our dataset consists of $29$ recordings from $9$ distinct subjects who are
researchers working in BigO. Subjects are aged between $25$ and $40$ years old,
normal sleepers without any sleep disorder. Six of them are male and three are
female. We requested them to keep information of their activity throughout the
recording sessions. The formulation of ground truth information is the
following: Every subject had to divide the recording session in sequential time
intervals, each being labeled as one of four possible states.

The four states are:

\begin{itemize}
\item Awake: subject wears her/his smartwatch and is not sleeping nor lying on
  her/his bed
\item In Bed: subject wears her/his smartwatch and is lying on her/his bed
\item Sleep: subject is sleeping
\item No Wear: subject does not wear her/his smartwatch
\end{itemize}

Annotation of recording session follows three rules:

\begin{enumerate}
\item Every recording session starts with a ``Recording Start'' event and ends
  with a ``Recording End'' event. After the Recording Start event, subject
  enters the Awake state.
\item The sequence of events [In Bed, Sleep Start, Sleep End, Off Bed] describes
  a sleeping period. Inside time interval [Sleep Start, Sleep End] subject is in
  Sleep state.
\item The sequence of events: [No wear, Wear] represents a non-wearing period,
  in which subject is in No Wear state.
\end{enumerate}

Recordings last for an average of $14.12$ hours and the mean duration of
sleeping period is $6.8$ hours. Two recordings contain two distinct sleeping
periods.

In order to evaluate the performance of the two methods we define the following
metrics. Each recording contains one or more sleep sessions. Let denote as
$X_{i}, i\in\{1, \ldots, 29\}$ every recording and $X_{i}^{j}\in\{i, \ldots\}$ the
$j$-th sleep session of $X_{i}$ recording. $S_{i}$ is the number of sessions
that recording $X_{i}$ contains and $\dot{S}_{i}$ the number of sessions that our
algorithm predicted.

Our primary goal is predicting correctly the number of sleep sessions for each
recording. We define the metric percentage of correct sleep sessions (CSS) as:
\begin{equation}
  \label{eq:6}
  CSS=\frac{\sum_{i=1}^{N}1,\text{ if } S_{i}=\dot{S}_{i}}{N}
\end{equation}
CSS is crucial since all sleep indicators metrics depend on it. 

If our algorithm is able to correctly detect most sleep sessions, we then define
another set of metrics that evaluates sleep indicators. For each
$X_{i}^{j}, i \in \{1, \ldots, N\}, j \in \{1, \ldots\}$ we define a set of variables:
\begin{itemize}
\item Sleep Start ($SS_{i}^{j}$) is the date-time in which sleep session $j$ of
  subject $i$ begins
\item Sleep End ($SE_{i}^{j}$) is the date-time in which sleep session $j$ of
  subject $i$ ends
\item Gross Sleep Time ($GST_{i}^{j}$) is the duration of sleep session $j$ of
  subject $i$: $GST_{i}^{j}=SE_{i}^{j}-SS_{i}^{j}$
\item Number of interrupts ($NI_{i}^{j}$) is the number of interrupts during
  sleep session $j$ of subject $i$
\item Total time of interrupts ($TTI_{i}^{j}$) is the aggregated duration of all
  interrupts during sleep session $j$ of subject $i$
\item Net Sleep Time ($NST_{i}^{j}$) is the duration of sleep session $j$ of
  subject $i$, without the interrupts: $NST_{i}^{j}=GST_{i}^{j}-TTI_{i}^{j}$
\end{itemize}

Evaluation of above variables is accomplished by computing the absolute error
(AE) in minutes, between prediction and ground truth for each recording.


On our BigO dataset, we evaluate metrics: $CSS$, $\mu(GST)$, $\mu(SS)$, and
$\mu(SE)$ were $\mu$ is the mean absolute error in minutes. It is not feasible
to validate algorithm’s performance on $NI$, $TTI$, and $NST$ due to the
absence of ground truth information on subject’s behavior throughout sleep
session.


\begin{table}
  \centering
  \caption{Evaluation of Sahe's algorithm on the BigO and Newcastle datasets. We
    present the mean ($\mu$), standard deviations ($\sigma$), and maximum
    ($\max{}$) absolute error (in minutes) per indicator; we also present the
    correct sleep sessions (CSS) per dataset.}
  \label{tab:sleep_results}
  \begin{tabular}{lcccccc}
    \toprule
    & \multicolumn{3}{c}{\textbf{BigO dataset}} & \multicolumn{3}{c}{\textbf{Newcastle dataset}} \\
    & $\mu$ & $\sigma$ & $\max{}$ & $\mu$ & $\sigma$ & $\max{}$ \\
    \midrule
    GST & $11.8$ & $10.1$ & $50.0$ & $17.9$ & $11.8$ & $44.0$ \\
    SS  & $ 9.3$ & $10.4$ & $43.6$ & $18.7$ & $19.7$ & $57.6$ \\
    SE  & $ 4.5$ & $ 3.3$ & $14.4$ & $11.6$ & $ 3.2$ & $15.9$ \\
    TTI & - & - & - & $16.7$ & $ 5.4$ & $28.0$ \\
    NI  & - & - & - & $ 9.9$ & $ 5.6$ & $19.0$ \\
    NST & - & - & - & $20.3$ & $10.0$ & $30.0$ \\
    \midrule
    $CSS$       & \multicolumn{3}{c}{$96.55\%$} & \multicolumn{3}{c}{$100\%$} \\
    \bottomrule
  \end{tabular}
\end{table}

\section{Real-world uses}
\label{sec:real_world}

An implementation of the proposed system was used in the context of the BigO
project in $33$ schools in Thessaloniki, Athens, Stockholm, Uppsala, and in two
childhood obesity clinics in Athens and Dublin. The system was supported by a a
number of front-end applications (portals), including a school portal, a
clinical portal, as well as portal for policy makers \cite{filos}. A total of
$3,700$ children contributed data, leading to the collection of $7,499$ days of
accelerometry, $4,692$ days of GPS, $169,316$ meal pictures and $35,012$ answers
to mood questions. Further details about the deployment of the system can be
found in \cite{adele, diou2020bigo, filos}. In \cite{filos}, the BigO system was
used to understand the environment close to the school in Stockholm, Sweden and
Thessaloniki, Greece. Results of \cite{filos} show that increased availability
of food or physical-activity related places near to school affects the students
behavior: students tend to eat more on food-related places if they are available
close to their school, but also tend to have increased physical activity levels
if there are more sports facilities available to them nearby.

As a public-health tool, BigO was used in three separate cases: (a) in
Stockholm, Sweden, to understand eating-location preference (school vs. retail
shops) in high-school populations, (b) in Thessaloniki, Greece, to compare
consumption of processed and ultra-processed foods in two
socioeconomically-distinct areas, and (c) in Stockholm, Sweden and in Larissa
and Thessaloniki, Greece, to monitor longitudinal population-behavior change
before and during the COVID-19 crisis. Ethical approval has been obtained in
Sweden, Stockholm (DNR: 2016/598-31) and in AUTH, Thessaloniki, Greece (DNR:
132649/2017 and 104673/2018)

As a clinical tool, the system was used in a population of overweight and obese
adolescents in the clinic for the Prevention and Management of Overweight and
Obesity, Division of Endocrinology, Metabolism and Diabetes, First Department of
Pediatrics, National and Kapodistrian University of Athens Medical School,
``Aghia Sophia'' Children’s Hospital, Athens, Greece. The system was used to
evaluate the correlation between BMI and behaviors measured by the BigO system,
as well as the effect of the COVID outbreak and subsequent measures on the
population behavior \cite{kassari2021evaluation1, kassari2021evaluation2}. In
particular, with the BigO system used as an intervention in a population of
$251$ obese or overweight children and adolescents, BMI decreased by $1.4\%$
($28.1 kg/m^{2}$ vs. $27.6 kg/m^{2}$, $p<0.001$) in all subjects
\cite{kassari2021evaluation1}.

Additionally, in \cite{nu13030880}, the relation between fast eating and BMI was
examined. In a Swedish population of $116$ students, those with higher BMI also
had faster eating rate ($+7.7g/min$ or $+27\%$, $p=0.012$). And in a second
population that included $748$ Swedish and $1084$ Greek high-school students,
similar observations were made. The Ph.D. thesis of Peter Fagerberg \cite{peter}
expanded on these results.

In \cite{nu15102321}, a population of $647$ students in Greece and Sweden was
analyzed with regards to their consumption of ultra-processed foods as well as
fruits and vegetables, before and after the COVID-19 pandemic. Analysis showed
that consumption of ultra-processed foods decreased during the pandemic while
consumption of fruits and vegetables increased.


Furthermore, a version of the proposed system was used by Eurostat in the
NTTS-2019 Big Data hackathon \cite{diou2019JIAOS}. More specifically, subjects
from around Europe (personnel of European National Statistics Offices) used the
system for approximately one month. This led to a dataset of $110$ users where
each user contributed data for $19$ days. This dataset was used in the hackathon
for the development of use-cases for use of Big Data for official statistics. An
example of results from participants using the collected data are available
online \cite{poland_eu_bd_hackathon}, demonstrating results from the Polish
Statistics Office (i.e., the winning team).


Finally, collaborating nutritionists are currently analyzing additional data of
school populations for the areas of Thessaloniki, Greece and Stockholm, Sweden.

\section{Conclusions}
\label{sec:conclusions}

We have presented the architecture and main behavioral indicator extraction
algorithms for a system that provides population-level measurements of behavior
and environment indicators, focusing on obesity prevention at the policy
level. The system relies on multi-modal information provided by smartphones and
smartwatches to continuously extract behavioral indicators, which can then be
aggregated at the individual and geographical area level. This aggregated
information, along with self-reported data from questionnaires and pictures can
then be used to support various applications in the school and clinical setting,
as well as for policy decision support.

The proposed system uses several signal processing algorithms to extract
indicators related to physical activity, types of locations visited,
transportation modes used, and sleep. The algorithms have been adapted for use
with modern smartphone and smartwatch devices, have been evaluated in datasets
available in the literature and demonstrate satisfactory effectiveness.

From a technical implementation viewpoint, one important challenge involves the
use of non-specialized equipment for data collection. Relying on the users'
personal devices for data collection places several limitations on the
parameters related to the acquired signals. These have to do with constraints on
the available signals imposed by the manufacturers, as well as limitations on
``background'' data collection, which require users to select complex
application permissions in the smartphone's OS. Overall, these obstacles can
lead to missing data and/or low compliance, however they can be overcome with
the collaboration of manufacturers in the future (e.g., to support systems for
participatory science and for monitoring population health indicators).

Another challenge involves the collection and analysis of large volumes of
sensor streams, as the number of concurrent users grows. The use of binary blobs
in CassandraDB by our system proved effective and can be scalable, since
CassandraDB can support distributed storage and, with the aid of Apache Spark,
distributed processing. Other solutions (e.g., TimescaleDB, InfluxDB) have been
recently proposed as mature implementations and they are worth
evaluating. Moreover, it is worth emphasizing that in all cases it is preferable
to process the raw sensor signals at the edge device, whenever possible, since
this limits the amount of information stored centrally, leading to computational
and privacy protection benefits.

The proposed system was used in studies for data collection by adults for the
NTTS2019 Big Data Hackathon and by children and adolescents in the context of
the BigO project. It has demonstrated its effectiveness in these real-world
setting, and shows that this participatory science approach can provide rich,
passively collected information to help support large-scale studies on complex
health challenges, such as obesity.

Finally, the proposed system is currently extended and adapted for the on-going,
EU-funded REBECCA project \cite{rebecca, kiriakidou2023integrating}.

\bibliographystyle{template/IEEEtran}
\bibliography{paper}

\end{document}